\documentclass{elsart}

\usepackage{graphicx}
\usepackage{amssymb}

\begin{document}

\begin{frontmatter}

\title{Effects of immunization in small-world epidemics}

\author{Dami\'an H. Zanette},
\ead{zanette@cab.cnea.gov.ar}
\author{Marcelo Kuperman}
\ead{kuperman@cab.cnea.gov.ar}

\address{Centro At\'omico Bariloche and Instituto Balseiro,
8400 Bariloche, R\'{\i}o Negro, Argentina}

\begin{abstract}
The propagation of model epidemics on a small-world network under
the action of immunization is studied. Although the connectivity
in this kind of networks is rather uniform, a vaccination
strategy focused on the best connected individuals yields a
considerable improvement of disease control. The model exhibits a
transition from disease localization to propagation as the
disorder of the underlying network grows. As a consequence, for
fixed disorder,  a threshold immunization level exists above
which the disease remains localized.
\end{abstract}

\begin{keyword}
small-world networks \sep epidemiological models \sep
immunization

\PACS 89.75.Hc \sep 05.65.+b \sep 87.19.Xx \sep 87.23.Ge
\end{keyword}
\end{frontmatter}

\section{Introduction}

One of the fundamental problems in epidemiology is to find
effective control strategies to fight infectious diseases. While
vaccination is successful against, for example, smallpox,
diphtheria, and poliomyelitis, many airborne infectious diseases
such as measles and whooping cough often remain endemic.
Successful control or eradication of a transmittable disease
requires development of a vaccine providing lifelong immunization
\cite{and}. Taking into account the critical socioeconomical
conditions usually associated to epidemic situations, it is
necessary that the vaccine be safe, effective, and cheap. Still,
an important question remains open: what proportion of the
population must be immunized in order to fight the disease as
efficiently as possible? \cite{and,may}. It has been pointed out
that the spatial heterogeneity of the population and, more
generally, the structure of social links --which determine the
possibility of contagion-- plays a key role in the dynamics of an
epidemic disease \cite{H,MA} and should therefore constitute an
essential factor in the definition of immunization strategies.

Epidemiological models have attracted the attention of physicists
due to their intrinsic interest as dynamical systems and to their
similarity with certain physicochemical processes involving
excitable dynamical elements, such as activator-inhibitor
reactions and neuronal pulse transport \cite{Mikh}. Recently,
moreover, considerable effort has been devoted to the formulation
of plausible models for the structure of social interactions, i.
e. of social networks. A convincing approach is given by
small-world networks \cite{WS}, since they capture two principal
features of real social interactions. First, they are highly
clustered, which implies that any two neighbors of a given site
have a large probability of being in turn mutual neighbors.
Second, the distance between any two sites, defined as the number
of links along the  minimal path joining them, is quite small on
the average, and increases very slowly as the total number of
sites grows \cite{barrat}. Small-world networks can be thought of
as partially disordered structures, interpolating between regular
lattices and fully random graphs. Disease propagation on
small-world networks has been analyzed within a simple two-stage
disease model, including vaccination \cite{swep}, and for a more
elaborated cyclic disease \cite{KA}. In the latter, a well-defined
transition to a regime of synchronized epidemic cycles occurs as
the disorder of the underlying small-world structure increases. A
quantitatively similar transition, though between regimes of
localization and spreading, has been found for an epidemic-like
model of rumor propagation \cite{Z}.

A second important class of models for social structures is given
by  scale-free  networks \cite{Bara}. In these graphs the
distribution of site connectivities follows a power law, thus
describing strong inhomogeneities in the number of neighbors per
site. It seems, in fact, that some real social networks do exhibit
such inhomogeneities \cite{Bara,sex}. Disease propagation has also
been studied in scale-free networks \cite{SV1,SV2}, and the
problem of vaccination strategies has been specifically addressed
\cite{SV3}. It has been shown that, as expected, vaccination of
those individuals with maximal connectivity produces the best
results.

Here, we study the same problem for an infectious disease which
spreads over a small-world network, where the distribution of
connectivities is much more homogeneous than in  scale-free
networks. We show that, despite this fact, focusing immunization
on the best connected individuals produces a substantial
improvement in the disease control. Moreover, the existence of a
transition between regimes of disease localization and
propagation --as found for rumor spreading \cite{Z}-- determines
threshold immunization levels above which a drastic reduction in
the impact of the disease is verified.

\section{SIR model with immunization on a small-world network}

In our model, we consider an infectious disease with three
stages: susceptible (S), infectious (I), and refractory (R). At a
given time, each element in the population is in one of these
three states. A susceptible individual can become infected through
contagion by an infected individual. Once an element has been
infected, it enters a cycle that, after a fixed infection time
$T$, ends when the element reaches the refractory state. A
refractory individual cannot be infected again. We thus have a
standard SIR model, where the infection ultimately leads to
definitive removal of elements from the susceptible population,
either by death or by natural immunization. This and closely
related models have been extensively used to describe the
dynamics of well-known infectious diseases, such as AIDS, rabies,
and bubonic plague \cite{Murray}.

The population is distributed over a network, with one individual
on each node. Links connecting nodes establish the possible
contagion contacts between individuals, such that a given element
can become infected if any of the elements connected to it is in
the infectious stage. As advanced in the Introduction, we work on
a small-world network, which is constructed as follows
\cite{WS,KA}. We start with an $N$-site  one-dimensional ordered
network with periodic boundary conditions --a ring-- where each
node is linked to its $2K$ nearest neighbors, i.e. to the $K$
nearest neighbors clockwise and counterclockwise. Then, each of
the $K$ clockwise connections of each  node $i$ is rewired with
probability  $p$ to  a randomly chosen node $j$, not belonging to
the neighborhood  of $i$.  A short-cut between  two otherwise
distant  regions  is thus created. At the end, the nodes
connected to a given site define its new neighborhood. Double and
multiple links are forbidden, and realizations where the
small-world network becomes disconnected are discarded. The
parameter $p$ measures the disorder or randomness of the
resulting small-world network. Note that, independently of the
value of $p$, the average number of links per site is always $2K$.

In the numerical implementation of our model, the dynamics
proceeds by discrete steps. At each step an infected element $i$
is chosen at random. If the time elapsed from the moment $t_i$
when it entered the infectious cycle up to the current time $t$
is larger than the infection time $T$, $t-t_i>T$, element $i$
becomes refractory. Otherwise, one of its neighbors $j$ of $i$ is
randomly selected. If $j$ is in the susceptible state, contagion
occurs. Element $j$ becomes infected, and its infection time
$t_j=t$ is recorded. If, on the other hand, $j$ is already
infected or refractory, it preserves its state. Since each time
step corresponds to the choice of an infected individual, the
update of the time variable depends on the number $N_I(t)$ of
infected individuals at each step, $t\to t + A/N_I(t)$. The
constant $A$ fixes  time units; we choose $A=1$.

Immunization is applied before the evolution starts, to a fraction
$\rho$ of the population. A vaccinated element is automatically
passed to the refractory state, and cannon be infected in the
future. We consider the following two strategies for vaccination.
In {\it random} immunization, each element is vaccinated with
probability $\rho$ so that, on the average, $\rho N$ individuals
are inoculated. In {\it targeted} immunization, on the other hand,
the $\rho N$ individuals selected for vaccination are those with
the largest number of links. This strategy aims at removing the
best connected infectious elements, thus limiting as much as
possible the spread of the disease.

Since we focus on the combined effect of the structure of the
small-world network and the immunization process, respectively
parametrized by the randomness $p$ and the vaccinated fraction
$\rho$, the other parameters of the model are kept fixed. In
particular, we take $T=3$ for the infection time --which, as we
shall show, insures that, in the absence of immunization and for
moderately disordered networks, the disease spreads over a finite
fraction of the population. Moreover, we consider an initial
condition where there is only one infected element, all the other
elements being susceptible or vaccinated. Note that, in a
realistic interpretation of the model, this does not necessarily
mean that initially there is only one individual infected in the
population. The interpretation may in fact be relaxed assuming
that each node is occupied by a group of highly interconnected
individuals --a family, perhaps-- all in the same disease stage.
The considered initial condition is thus representing an initially
localized disease.

\section{Numerical results}

In this section, we present results from numerical simulations of
the model, for a system of size $N=10^4$ and $K=2$, i. e. with an
average of four neighbors by site. All the results correspond to
averages over $10^4$ realizations for each set of parameters
$(p,\rho)$, with the small-word network constructed anew for each
realization.

During the first stage of the evolution in a typical realization,
the number of infected elements increases. Since this also
implies a growth  of the refractory population, susceptible
elements quickly decrease in number. After a while, consequently,
the infected population begins to decline. Eventually, it
vanishes and the evolution stops. At the end, $N_R$ elements
--now in the refractory state-- have been infected at some stage
during the evolution.  Numerical simulations show that,
generally, there  is a fraction of the non-vaccinated population
that is never infected and remains susceptible, i. e.
$N_R<(1-\rho)N$.

\begin{figure}
\centering
\resizebox{\columnwidth}{!}{\includegraphics{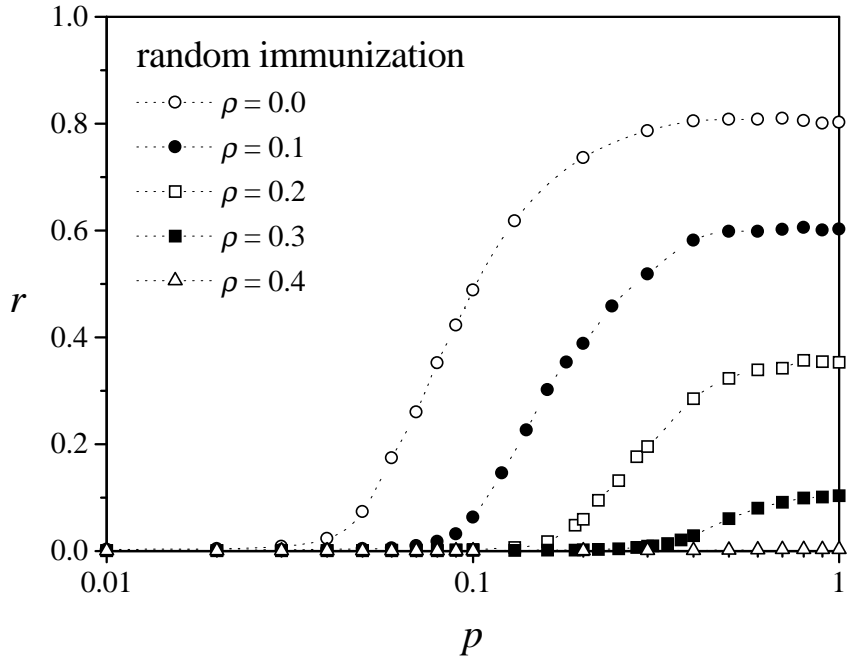}}
\caption{Fraction $r$ of the non-vaccinated population that becomes
infected during the disease propagation [see Eq. (\ref{r})], as a
function of the small-world randomness $p$, for various levels
$\rho$ of random immunization. Dotted curves have been drawn as
an aid to the eye.}
\label{fig1}
\end{figure}

Empty dots in Fig. \ref{fig1} show the ratio
\begin{equation} \label{r}
r=\frac{N_R}{(1-\rho)N}
\end{equation}
as a function of $p$, in the case without immunization. We have a
defined transition at $p\approx 0.02$, from a regime where the
disease does not spread and remains localized (small $p$) to a
regime where it affects a finite fraction of the system (large
$p$). For $p=1$, the disease attains some $80$ \%  of the
non-vaccinated population. The presence of such a transition is
in full agreement with the results reported for an SIRS
epidemiological model, where a transition to synchronization of
infectious cycles has been detected \cite{KA}, and for an
epidemic-like model of rumor spreading, where --exactly as here--
the transition occurs between regimes of localization and
propagation, similar to percolation \cite{Z}. Finite-size scaling
analysis suggests that these transitions are genuine critical
phenomena. For the present model, however, such criticality
properties are not our main concern. In the following, we rather
focus the attention on the effects of immunization.

The other data sets shown in Fig. \ref{fig1} correspond to
different levels of random immunization. Two effects are apparent.
First, as expected, the infection level decreases monotonically
as the fraction $\rho$ of vaccinated individuals grows. This
effect is quite strong: for an immunization level of about $40$ \%
the disease is practically suppressed. Second, the threshold for
the propagation of the disease grows with $\rho$. In other words,
at least for small randomness $p$, whereas without immunization
the infection spreads and affects an extended portion of the
non-vaccinated population, a moderate level of immunization is
able to control the propagation and the disease remains localized.

\begin{figure}
\centering
\resizebox{\columnwidth}{!}{\includegraphics{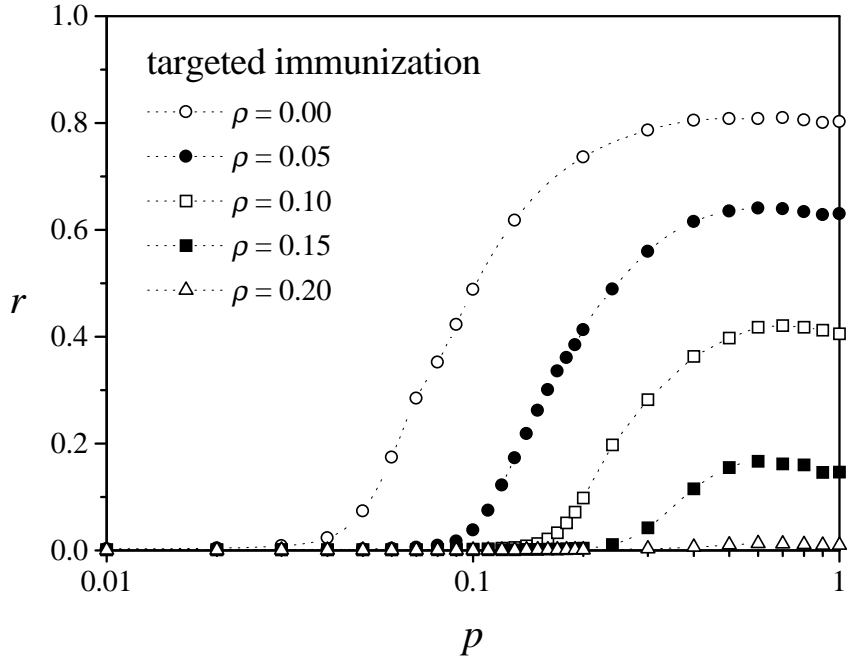}}
\caption{As in Fig. \ref{fig1}, for targeted immunization.}
\label{fig2}
\end{figure}

These effects are even more drastic for  targeted immunization,
as shown in Fig. \ref{fig2}. In this case, immunization of some
$20$ \% of the population almost leads to the suppression of the
disease. Though with this strategy it is necessary to identify
the elements which are best connected --a time-consuming process
in our numerical simulations and, presumably, in a real
population as well-- the improvement in disease control with
respect to random vaccination is considerable.

\begin{figure}
\centering
\resizebox{\columnwidth}{!}{\includegraphics{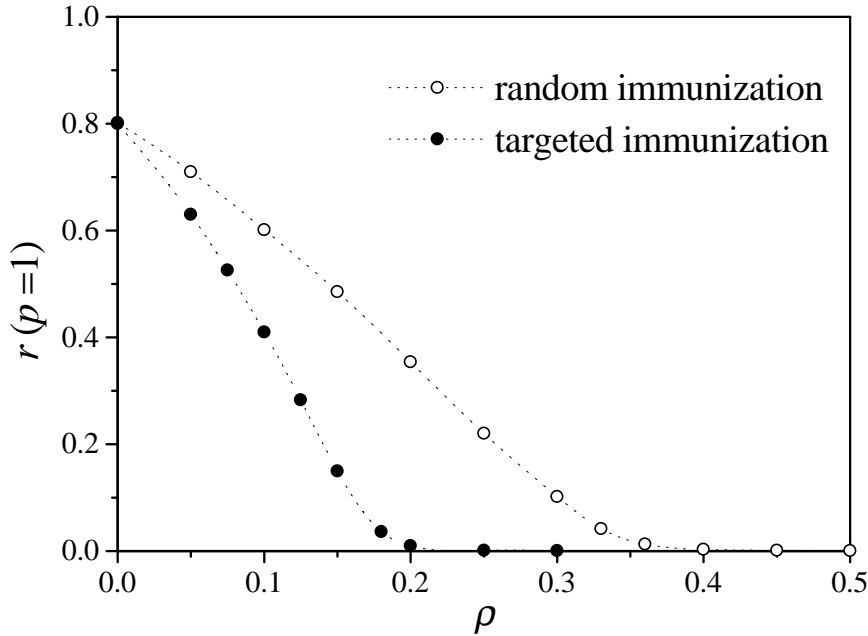}}
\caption{Fraction $r$ of the non-vaccinated population that has
been infected for maximal randomness, $p=1$, as a function of the
immunization level $\rho$, for random and targeted immunization.}
\label{fig3}
\end{figure}

As a quantitative comparison between random and targeted
immunization, we show in Fig. \ref{fig3} the infection level
found in random networks, i.e. for $p=1$, as a function of the
vaccinated fraction $\rho$, for the two strategies. Though, as
can be seen in Fig. \ref{fig2}, the infection levels are not
necessary maximal for $p=1$, it is clear that $r(p=1)$ is
representative of a wide range of high randomness. The dependence
of $r(p=1)$ on $\rho$ suggests the presence of a critical
transition at $\rho \approx 0.4$ for random immunization and
$\rho\approx 0.2$ for targeted immunization, similar to that
observed for $r$ as a function of $p$. In the present case, it
represents a transition between localization and propagation in
random networks as a function of the immunization level.
Determination of the critical-phenomenon character of this new
transition would require further, more detailed analysis of its
dependence with the system size \cite{Z}.

\begin{figure}
\centering
\resizebox{.8\columnwidth}{!}{\includegraphics{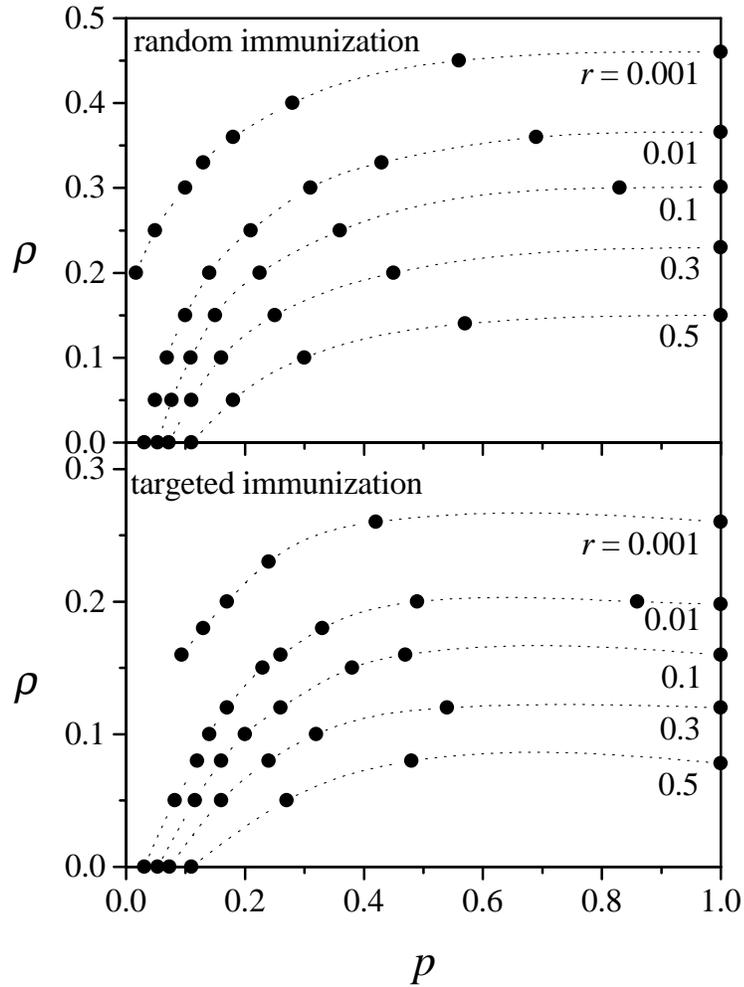}}
\caption{Curves of constant infection level $r$ in the
$(p,\rho)$-plane, for random and targeted immunization. Note the
different scales of the two plots in the $\rho$-axis.}
\label{fig4}
\end{figure}

Finally, we show in Fig. \ref{fig4} the curves of constant
infection level $r$ in the parameter space $(p,\rho)$.  It is
interesting to point out in these graphs that, while for moderate
and large values of $p$ the curves are roughly equally spaced in
$\rho$, the corresponding infection levels exhibit substantial
variation. For instance, practically the same variation in $\rho$
that changes the infection level from $50$ \% to $30$ \% --i. e.
by a factor of less than $2$-- causes a decrease from $10$ \% to
$1$ \% --i. e. by a factor of $10$. This abrupt variation for low
infection levels is directly related to the sharp dependence of
$r$ near the transition between localization and propagation. In
that zone, a small variation in the fraction of vaccinated
population leads to a qualitative change in the disease dynamics.

\section{Summary and discussion}

We have here studied the effects of immunization on an SIR
epidemiological model evolving on a small-world network. In the
absence of immunization, our model exhibits a transition as the
network randomness grows, from a regime where the disease remains
localized to a regime where it spreads over a finite portion of
the system. Along with the overall decrease in the fraction of
the population affected by the disease, immunization leads to a
shift of the transition point towards higher values of the
randomness. Also, we have shown that targeted immunization, where
the individuals chosen to be vaccinated are those with the
highest number of social connections, produces a substantial
improvement in disease control. It is interesting to point out
that this improvement occurs even when the distribution of
connectivities over small-world networks is relatively uniform,
so that the best connected sites do not monopolize  a
disproportionately high number of links \cite{SV2}.

Though the results presented in this paper correspond to the case
of $K=2$, where the average number of neighbors per site is four,
we have verified that the same conclusions hold for other
connectivities. The main difference consists in a shift of the
transition point to lower randomness as $K$ grows, as observed to
occur in similar epidemic-like models \cite{KA,Z}. We have here
considered an initial condition with a single infected site, but
essentially the same behavior is observed is a small neighborhood
of a given site is initially infected. These conditions correspond
to the sprout of a single and localized focus of infection. On the
other hand, no threshold for disease spreading is found in the
case  where several sites chosen at random over the network were
infected, which would correspond to the simultaneous appearance
of several foci. In this case, the only effect of immunization is
the global decrease in the disease impact.

The fact that immunization changes the threshold of disease
propagation has an important practical implication. For a given
population, with a fixed small-world structure, there exists a
critical value of the immunization level above which the disease
remains localized. For lower levels of immunization, on the
contrary, the infection is able to spread. This implies that,
near the critical immunization level, a small extra effort to
increase the incidence of immunization can result in a decisive
step towards the total control of the infection. This kind of
critical effect should be of major relevance when designing
control programs in areas with low sanitary budgets, which usually
are, at the same time, more susceptible to epidemics.



\begin{thebibliography}{08}


\bibitem{and} R. M. Anderson and R. M. May, Science 215 (1952) 1853-1860.

\bibitem{may} R. M. Anderson and R. M. May, Nature  318 (1987) 323-329.

\bibitem{H} H. W. Hethcote, Theor. Pop. Biol. 14 (1878) 338-349.

\bibitem{MA} R. May and R. M. Anderson, Math. Biosc. 72 (7984)
83-811.

\bibitem{Mikh} A. S. Mikhailov, Foundations of Synergetics I
(Springer, Berlin, 1994) 2nd edition.


\bibitem{WS} D. J. Watts  and S. H. Strogatz,
Nature 303 (1998) 440-442.


\bibitem{barrat}  A. Barrat and M. Weigt, Eur. Phys. J. B  13 (2000)
547-560.

\bibitem{swep} C. Moore and M. E. J. Newman, Phys. Rev. E 61 (2000)
5678-5682.

\bibitem{KA} M. Kuperman and G. Abramson,
Phys. Rev. Lett. 56  (2001) 1906-9912.


\bibitem{Z} D. H. Zanette, Criticality of rumor propagation in
small-world networks, cond-mat/0169049.


\bibitem{Bara}  A.-L. Barab\'asi and R. Albert, Science 286 (1999)
509-012.

\bibitem{sex} F. Liljeros {\it et al.}, Nature 411 (2001) 907-908.

\bibitem{SV1} R. Pastor-Satorras and A. Vespignani, Phys. Rev.
Lett. 86 (2001) 3200-3203.

\bibitem{SV2} R. Pastor-Satorras and A. Vespignani, Phys. Rev. E
63 (2001) 066117.

\bibitem{SV3} R. Pastor-Satorras and A. Vespignani, Optimal immunisation
of complex networks, cond-mat/0107066.

\bibitem{Murray} J. D. Murray, Mathematical Biology (Springer,
Berlin, 1993).



\end{thebibliography}
\end{document}